\documentclass[aps,preprint,groupedaddress,showpacs,floatfix]{revtex4}
\usepackage{graphicx}

\begin{document}

\title{Light scalar mesons and two-kaon correlation functions}
\author {
N.N. Achasov$^{\,a}$ \email{achasov@math.nsc.ru} and A.V.
Kiselev$^{\,a,b}$, \email{kiselev@math.nsc.ru}}

\affiliation{$^a$Laboratory of Theoretical Physics,
 Sobolev Institute for Mathematics, 630090, Novosibirsk, Russia\\
$^b$Novosibirsk State University, 630090, Novosibirsk, Russia}

\date{\today}

\begin{abstract}

It is shown that the recent data on the $K^0_S K^+$ correlation in
Pb-Pb interactions agree with the data on the
$\gamma\gamma\to\eta\pi^0$ and $\phi\to\eta\pi^0\gamma$ reactions
and support the four-quark model of the $a_0(980)$ meson. It is
shown that the data does not contradict the validity of the
Gaussian assumption. The study of two-kaon correlations could
provide more information about light scalar mesons after
increasing the accuracy of the experimental and theoretical
descriptions.
\end{abstract}
 \maketitle

\section{Introduction}

The $a_0(980)$ and $f_0(980)$ mesons are well-established parts of
the proposed light scalar meson nonet \cite{pdg-2016}. From the
beginning, the $a_0(980)$ and $f_0(980)$ mesons became one of the
central problems of nonperturbative QCD, as they are important for
understanding the way chiral symmetry is realized in the
low-energy region and, consequently, for understanding
confinement. Many experimental and theoretical papers have been
devoted to this subject.

There is much evidence that supports the four-quark model of light
scalar mesons \cite{jaffe}.

First, the suppression of the $a_0(980)$ and $f_0(980)$ resonances
in the $\gamma\gamma\to\eta\pi^0$ and $\gamma\gamma\to\pi\pi$
reactions, respectively, was predicted in 1982 \cite{fourQuarkGG},
$\Gamma_{a_0\gamma\gamma}\approx \Gamma_{f_0\gamma\gamma}\approx
0.27$ keV, and confirmed by experiment \cite{pdg-2016}. The
elucidation of the mechanisms of the $\sigma(600)$, $f_0(980)$,
and $a_0(980)$ resonance production in $\gamma\gamma$ collisions
confirmed their four-quark structure \cite{AS88,annsgnGamGam}.
Light scalar mesons are produced in $\gamma\gamma$ collisions
mainly via rescatterings, that is, via the four-quark transitions.
As for $a_2(1320)$ and $f_2(1270)$ (the well-known $q\bar q$
states), they are produced mainly via the two-quark transitions
(direct couplings with $\gamma\gamma$).

Second, the argument in favor of the four-quark nature of
$a_0(980)$ and $f_0(980)$ is the fact that the $\phi(1020)\to
a_0\gamma$ and $\phi(1020)\to f_0\gamma$ decays go through the
kaon loop: $\phi\to K^+K^-\to a_0\gamma$, $\phi\to K^+K^-\to
f_0\gamma$, i.e., via the four-quark transition
\cite{achasov-89,achasov-97,a0f0,our_a0,achasov-03}. The kaon-loop
model was suggested in Ref. \cite{achasov-89} and confirmed by
experiment ten years later \cite{SNDa0,f0exp,kloea0}.

It was shown in Ref. \cite{achasov-03} that the production of
$a_0(980)$ and $f_0(980)$ in $\phi\to a_0\gamma\to
\eta\pi^0\gamma$ and $\phi\to f_0\gamma\to\pi^0\pi^0\gamma$ decays
is caused by the four-quark transitions, resulting in strong
restrictions on the large-$N_C$ expansions of the decay
amplitudes. The analysis showed that these constraints give new
evidence in favor of the four-quark nature of the $a_0(980)$ and
$f_0(980)$ mesons.

Third, in Refs. \cite{agsh,ak-07} it was shown that the
description of the $\phi\to K^+K^-\to\gamma a_0(980)/f_0(980)$
decays requires virtual momenta of $K (\bar K)$ greater than $2$
GeV, while in the case of loose molecules with a binding energy
about 20 MeV, they would have to be about 100 MeV. Besides, it
should be noted that the production of scalar mesons in the
pion-nucleon collisions with large momentum transfers also points
to their compactness \cite{ADS98}.

Fourth, the data on semileptonic $D^+_s\to s\bar s\, e^+\nu\to
[\sigma(600)+f_0(980)]e^+\nu\to \pi^+\pi^-e^+\nu$ decays are also
in favor of the four-quark nature of $\sigma(600)$ and $f_0(980)$
\cite{dsdecay}. Unfortunately, at the moment the statistics is
rather poor, and thus new high-statistics data are highly
desirable. No less interesting is the study of semileptonic decays
of $D^0$ and $D^+$ mesons -- $D^0\to d\bar u\, e^+\nu\to
a_0^-e^+\nu\to\pi^-\eta e^+\nu$, $D^+\to d\bar d\, e^+\nu\to a_0^0
e^+\nu\to\pi^0\eta e^+\nu$ (or the charged-conjugated ones) and
$D^+\to d\bar d\, e^+\nu\to [\sigma(600)+f_0(980)]e^+\nu\to
\pi^+\pi^-e^+\nu$ -- which have not been investigated yet
\cite{dsdecay,dsdecayConf}. It is very tempting to study light
scalar mesons in semileptonic decays of $B$ mesons
\cite{dsdecayConf}: $B^0\to d\bar u\, e^+\nu\to
a_0^-e^+\nu\to\pi^-\eta e^+\nu$, $B^+\to u\bar u\, e^+\nu\to a_0^0
e^+\nu\to\pi^0\eta e^+\nu$, $B^+\to u\bar u\, e^+\nu\to
[\sigma(600)+f_0(980)]e^+\nu\to \pi^+\pi^-e^+\nu$.

it was also shown in Refs. \cite{annshgn-94,annshgn-07} that the
linear $S_L(2)\times S_R(2)$ $\sigma$ model \cite{gellman}
reflects all of the main features of low-energy $\pi\pi\to\pi\pi$
and $\gamma\gamma\to\pi\pi$ reactions and agrees with the
four-quark nature of light scalar mesons. This allowed for the
development of a phenomenological model with the right analytical
properties in the complex $s$ plane that took into account the
linear $\sigma$ model and the background \cite{our_f0_2011}. This
background has a left cut inspired by crossing symmetry, and the
resulting amplitude agrees with results obtained using the chiral
expansion, dispersion relations, and the Roy equation
\cite{sigmaPole}, and with the four-quark nature of the
$\sigma(600)$ and $f_0(980)$ mesons as well.

Recently, the ALICE Collaboration's investigation of $K^0_S K^\pm$
correlation \cite{alice-2017} determined that $a_0(980)$ is a
four-quark state. This conclusion was made on the basis that
masses and coupling constants obtained in the four-quark-based
scenario from the data on $\phi\to\eta\pi^0\gamma$ and
$\phi\to\pi^0\pi^0\gamma$ decays \cite{a0f0,kloea0,our_a0}
accurately describe the data on two-kaon correlations, in
contradistinction to another set of parameters \cite{martin}.

Statistically significant data on two-kaon correlations appeared
recently. In 2006, the STAR Collaboration presented data on $K^0_S
K^0_S$ correlation in Au-Au interactions \cite{star-2006}. Both
$a^0_0(980)$ and $f_0(980)$ are created in the process.

Recently, the ALICE Collaboration published data on $K^0_S K^\pm$
correlations in Pb-Pb interactions \cite{alice-2017}, and
$a^\pm_0(980)$ is created in these reactions.

In 2015, the authors of \cite{aks-2015} presented an analysis on
the Belle data on the $\gamma\gamma\to \eta\pi^0$ reaction
together with KLOE data on the decay $\phi\to\eta\pi^0\gamma$.
Here, we present a new analysis which additionally includes the
ALICE data on $K^0_S K^+$ correlation \cite{alice-2017}.

We justify the $a_0(980)$ four-quark nature on a higher level than
that in Ref. \cite{alice-2017}: the set of different data (the
Belle data on $\gamma\gamma\to \eta\pi^0$, the KLOE data on
$\phi\to\eta\pi^0\gamma$, and the ALICE data on two-kaon
correlation) is simultaneously described in a scenario based on
the four-quark model \cite{jaffe}. In this scenario the coupling
constants obey (or almost obey) the relations \cite{achasov-89}

\begin{eqnarray} &
g_{a_0\eta\pi^0}=\sqrt{2}
\mbox{sin}(\theta_p+\theta_q)g_{a_0K^+K^-}=(0.85\div 0.98)
g_{a_0K^+K^-} , & \nonumber\\ & g_{a_0\eta'\pi^0}=-\sqrt{2}
\mbox{cos}(\theta_p+\theta_q)g_{a_0K^+K^-}=-(1.13\div 1.02)
g_{a_0K^+K^-} , & \label{FourQrelations}
\end{eqnarray}

\noindent and the coupling to the $\gamma\gamma$ channel is small.
Here $g_{a_0\eta\pi^0}=0.85\, g_{a_0K^+K^-}$ and
$g_{a_0\eta'\pi^0}=-1.13\, g_{a_0K^+K^-}$ for $\theta_p=-18^\circ$
and $g_{a_0\eta\pi^0}=0.98\, g_{a_0K^+K^-}$ and
$g_{a_0\eta'\pi^0}=-1.02\, g_{a_0K^+K^-}$ for
$\theta_p=-11^\circ$. The $\theta_q=54.74^\circ$.

Our description takes into account the $a_0'$ meson and uses
one-loop scalar propagators with good analytical properties; see
Sec. \ref{formRes}.

The approach is based on Ref. \cite{ledlub-1982}, which in turn
was based on the assumption of an ideal chaotic Gaussian source,
which requires that the correlation strength $\lambda$ be equal to
unity; for details, see Ref. \cite{lambdaClarification}. We show
that the data is described well with $\lambda=1$, what didn't
manage to be made in Ref. \cite{alice-2017}.

Note that we do not use the STAR and ALICE data on the correlation
of identical kaons
\cite{star-2006,aliceIdentical1,aliceIdentical2} because in the
charged case $a_0(980)$ is not created, and the neutral case deals
both with isospin I=0 and I=1, i.e., a similar simultaneous
analysis would require taking into account $f_0(980)$,
$f_2(1270)$, and $\sigma(600)$ and the reactions $\gamma\gamma\to
\pi^0\pi^0$, $\phi\to\pi^0\pi^0\gamma$, and $\pi\pi\to\pi\pi$.
This is a rather complicated problem, and we hope to return to it
in the future.

\section{Formalism and results}\label{formRes}

Let us briefly consider the formalism used in Ref.
\cite{alice-2017}, which is based on that in Ref.
\cite{ledlub-1982}. The scattering amplitude is (Eq. (6) of Ref.
\cite{alice-2017}):
\begin {equation} f(k^*)=\frac{\gamma_{a_0\to K\bar
K}}{m_{a_0}^2-s-i(\gamma_{a_0\to K\bar K}k^*+\gamma_{a_0\to
\pi\eta}k_{\pi\eta})}.\label {f}
\end{equation}

Here the denominator is the inverse propagator of $a^+_0$ in a
Flatt\'e-like form \cite{flatte}, $s$ is the invariant two-kaon
mass squared, $k^*$ is the kaon momentum in the kaon pair rest
frame,
\begin {equation}
k^*=\frac{\sqrt{(s-(m_{K^0_S}-m_{K^+})^2)(s-(m_{K^0_S}+m_{K^+})^2)}}{2\sqrt{s}},
\end{equation} \noindent and $k_{\pi\eta}$ is the corresponding
$\pi\eta$ momentum.

The correlation $C(k^*)$ is (Eq. (9) of Ref. \cite{alice-2017})
\begin {equation}
C(k^*)=1+\frac{\lambda}{2}\bigg(\frac{1}{2}\bigg|\frac{f(k^*)}{R}\bigg|^2+2\frac{Re
f(k^*)}{\sqrt{\pi}R}F_1(2k^*R)-\frac{Im
f(k^*)}{R}F_2(2k^*R)\bigg),\label{corrForm}
\end{equation} where $R$ is the radius parameter from the spherical Gaussian source distribution, $\lambda$ is the correlation
strength, and  \begin{equation}F_1(z)=\frac{e^{-z^2}}{z}\int_0^z
e^{x^2}dx;\hspace{5pt} F_2(z)=\frac{1-e^{-z^2}}{z}.\label{F12}
\end{equation}

The Flatt\'e propagator is not adequate for studying $f_0(980)$
and $a_0(980)$; see Refs.
\cite{achasov-1995,achasov-97,agsh,ourProp,AS2017}. As in Ref.
\cite{aks-2015}, we use one-loop propagators and take into account
the $a'^+_0$ meson, so Eq. (\ref{corrForm}) is modified:
\begin {equation} f(k^*)=\frac{2}{\sqrt{s}}\sum_{S,S'}\frac{g_{S
K^0_S K^+}G_{SS'}^{-1}g_{S'K^0_S K^+}}{16\pi}.
\end{equation}

\noindent where $S,S'=a^+_0,a'^+_0$, and the constants $g_{S K^0_S
K^+}=-g_{S K^0_L K^+}=g_{SK^+K^-}$. The matrix of the inverse
propagators is \begin{equation}G_{SS'}\equiv G_{SS'}(m)=\left(
\begin{array}{cc}
D_{a'_0}(m)&-\Pi_{a'_0a_0}(m)\\-\Pi_{a'_0a_0}(m)&D_{a_0}(m)\end{array}\right),\end{equation}
\begin{equation}\Pi_{a'_0a_0}(m)=\sum_{a,b} \frac{g_{a'_0 ab}}{g_{a_0
ab}}\Pi^{ab}_{a_0}(m)+C_{a'_0 a_0},\end{equation} \noindent where
$m=\sqrt{s}$, and the constant $C_{a_0'a_0}$ incorporates the
subtraction constant for the transition $a_0(980)\to(0^-0^-)\to
a_0'$ and effectively takes into account the contributions of
multiparticle intermediate states to the $a_0\leftrightarrow a_0'$
transition. The inverse propagator of the scalar meson $S$
\cite{achasov-89,achasov-97,aks-2015,adsh-79} is
\begin{equation} \label{propagator} D_S(m)=m_S^2-m^2+\sum_{ab}[Re
\Pi_S^{ab}(m_S^2)-\Pi_S^{ab}(m^2)],
\end{equation}
\noindent where $\sum_{ab}[Re \Pi_S^{ab}(m_S^2)-
\Pi_S^{ab}(m^2)]=Re\Pi_S(m_S^2)- \Pi_S(m^2)$ takes into account
the finite-width corrections of the resonance which are the
one-loop contributions to the self-energy of the $S$ resonance
from the two-particle intermediate  $ab$ states. We take into
account the intermediate states $\eta\pi^+,K\bar K$, and $
\eta'\pi^+$ in the $a^+_0(980)$ and $a'^+_0$ propagators:
\begin{equation}
\Pi_S=\Pi_S^{\eta\pi^+}+\Pi_S^{K^0_SK^+}+\Pi_S^{K^0_LK^+}+
\Pi_S^{\eta'\pi^+}.
\end{equation}

The forms of $\Pi_S^{ab}(m)$ are expressed in Appendix I.

Equipped with these formulas, we fit the "previous" data (i.e.,
the data on $\gamma\gamma\to \eta\pi^0$ \cite{uehara} and
$\phi\to\eta\pi^0\gamma$ \cite{kloea0} reactions) as in Ref.
\cite{aks-2015} simultaneously with the ALICE data on $K^0_SK^+$
correlation (29 points from the upper-left panel in Fig. 2 of Ref.
\cite{alice-2017}). Only statistical errors are taken into
account.

Unfortunately, the ALICE Collaboration did not publish the data in
the form of a table, with statistical, systematic, and total
errors for combined $K^0_S K^+$ and $K^0_S K^-$ data sets. For
safety, we neglect systematic error and do not fit the data on
$K^0_S K^-$ (the data sets are consistent).

We perform four analogs of Fit 1 of Ref. \cite{aks-2015}; see
Table I and Fig. \ref{corrPlot}. Parameters that are not mentioned
above are in Table II of Appendix II. To fit the "previous" data
we use the same $\chi^2$ functions with the same restrictions,
including fixing the $a'_0$ mass at $1400$ MeV and terms that
guarantee being close to the four-quark model relations
(\ref{FourQrelations}); for details, see Ref. \cite{aks-2015}. The
$\chi^2_{corr}$ in Table I is the usual $\chi^2$ function built on
the $K^0_S K^+$ correlation data.

In Table I, Fit 1 is for free $\lambda$ and $R$, and Fit 2 is for
$\lambda=1$. One can see that the quality of Fit 2 is also good so
the data does not contradict $\lambda$ being equal to unity.

\begin{center}
Table I. Properties of the resonances and the description quality.
\begin{tabular}{|c|c|c|c|c|c|}\hline

Fit & 1 & 2 & 3 & 4 \\ \hline

$m_{a_0}$, MeV & $995.1$ & $1003$ & $993.9$ & $993.9$ \\ \hline

$g_{a_0K^+K^-}$, GeV  & $2.70$ & $2.73$ & $2.75$ & $2.75$
\\ \hline

$g_{a_0 \eta\pi}$, GeV  & $2.85$ & $2.95$ & $2.74$ & $2.74$
\\ \hline

$g_{a_0 \eta'\pi}$, GeV  & $-2.79$ & $-2.81$ & $-2.86$ & $-2.86$
\\ \hline

$m_{a'_0}$, MeV & $1400$ & $1400$ & $1400$ & $1400$ \\ \hline

$g_{a_0' K^+K^-}$, GeV  & $0.87$ & $1.04$ & $1.63$ & $1.63$
\\ \hline

$g_{a_0' \eta\pi}$, GeV  & $-2.33$ & $-2.72$ & $-3.12$ & $-3.12$
\\ \hline

$g_{a_0' \eta'\pi}$, GeV  & $-6.73$ & $-6.56$ & $-4.75$ & $-4.75$
\\ \hline

$C_{a_0 a'_0}$, GeV$^2$ & $0.146$ & $0.133$ & $0.021$ & $0.021$
\\ \hline

$\lambda$ & $0.53$ & $1$ & $0.73$ & $1$
\\ \hline

$R$, fm & $5.0$ & $6.7$ & $5.6$ & $6.8$
\\ \hline

$\chi^2_{\gamma\gamma}$ / $36$ points & $13.1$ & $19.0$ & $12.4$ &
$12.4$
\\ \hline

$\chi^2_{sp}$ / $24$ points & $24.7$ & $25.6$ & $24.5$ & $24.5$
\\ \hline

$\chi^2_{corr}$ / $29$ points & $19.0$ & $28.2$ & $24.8$ & $40.4$
\\ \hline

($\chi^2_{\gamma\gamma}$+$\chi^2_{sp}$+$\chi^2_{corr}$)/n.d.f. &
$56.9/73$ & $72.8/74$ & $61.6/73$ & $77.2/74$
\\ \hline

\end{tabular}
\end{center}

Fits 3 and 4 are for the parameters of Fit 1 of Ref.
\cite{aks-2015}, with free $\lambda$ and $R$ and with $\lambda=1$,
respectively. One can see that Fit 3 describes the data quite
well, while Fit 4 does not describe the data on correlation with a
perfect $\chi^2$, though the description is not very bad since
errors are small and systematic errors are neglected.

\begin{figure}
\begin{center}
\begin{tabular}{ccc}
\includegraphics[width=8cm]{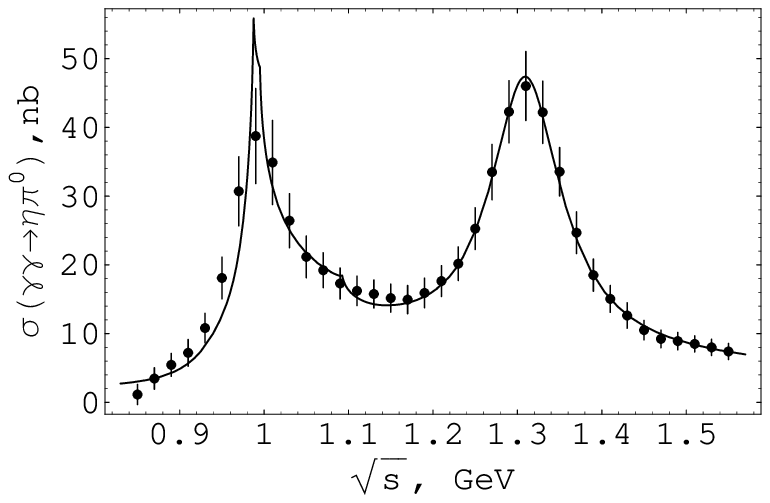}& \includegraphics[width=8cm]{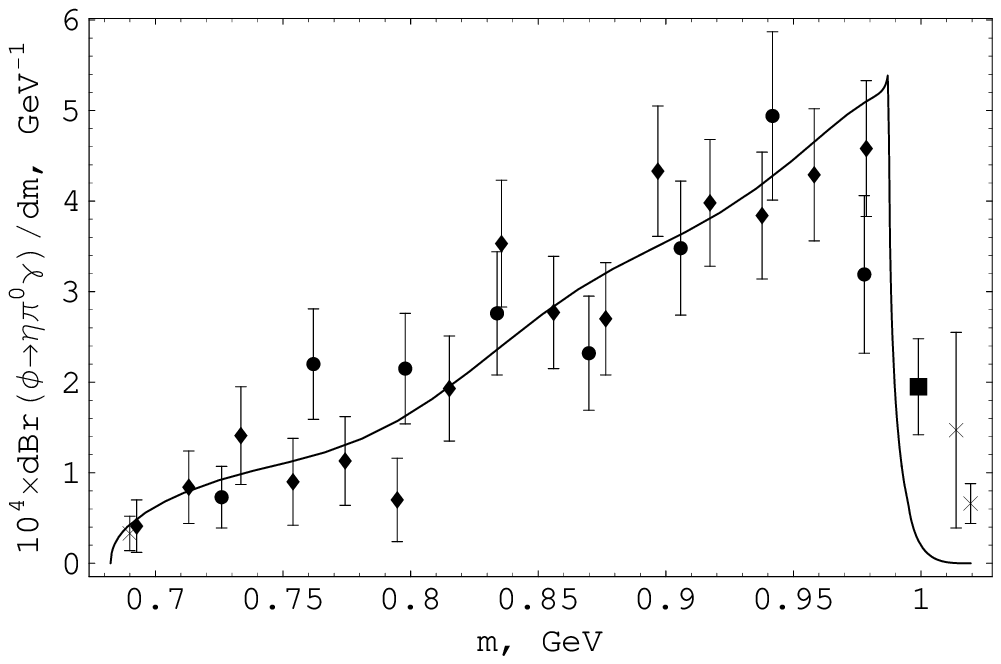}\\ (a)&(b)
\end{tabular}
\end{center}
\caption{(a) The $\gamma\gamma\to \eta\pi^0$ cross section, the
curve is Fit 1, and the data points are from Belle \cite{uehara}.
Note that the Belle data represent the averaged cross section
(each bin is 20 MeV). (b) Plot of the Fit 1 curve and the KLOE
data (points) \cite{kloea0} on the $\phi\to\eta\pi^0\gamma$ decay;
$m$ is the invariant $\eta\pi^0$ mass. Cross points are omitted in
the fitting. } \label{oldData}
\end{figure}

\begin{figure}
\centerline{\includegraphics[width=12cm]{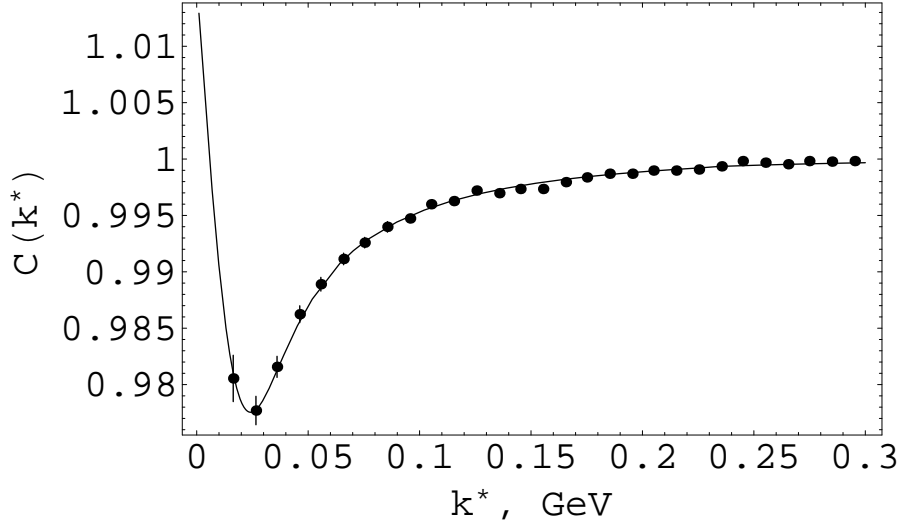}}
\caption {$K^0_SK^+$ correlation: the solid line corresponds to
Fit 1, and the points are experimental data \cite{alice-2017}.}
\label{corrPlot}\end{figure}

The difference in the $a_0$ features between Fits 1--2 and Fits
3--4 (with the "old" parameters) is rather small: the $a'_0$
features are more fluid, as was observed in Ref. \cite{aks-2015}.
The description of "previous" data for Fit 1 is shown in Fig.
\ref{oldData}: it is close to that in Ref. \cite{aks-2015}. The
correlation is shown in Fig. \ref{corrPlot}.

Analogs of other fits from Ref. \cite{aks-2015} could be obtained
in the same way.

As in Ref. \cite{aks-2015}, we do not calculate errors of the
parameters. In our case, the minimized function has more than one
minimum, for example, one with $\lambda=0.53$ (Fit 1) and another
with $\lambda=0.66$ and $m_{a_0}= 1012$ MeV. The last value
exceeds the usually obtained ones, but is also not excluded. The
values of the minimized function differ by less than 1 in these
minima, while for $\lambda$ in the intermediate region
$0.53$--$0.66$ the deviation from the minimum values is greater
than 1.

In Ref. \cite{alice-2017} the obtained values of $\lambda$ were
much less than 1 and not far from $\lambda=0.53$ in Fit 1. We are
able to obtain Fit 2 with $\lambda=1$ primarily because of the
presence of $a_0'$ and the fact that we vary the $a_0$ parameters.
$a_0'$ gives a notable contribution to the correlation: its
removal raises $\chi^2_{corr}$ from 19 to 57 in Fit 1, and from 28
to 107 in Fit 2, while the other parameters remain the same. Also,
we use a better propagator for the scalar particles.

Fits 1--3 show that the whole set of experimental data could be
described in the four-quark model of $a_0(980)$. Moreover, the
results of the previous analysis are well consistent with the data
on correlation.

The predictive power of the data on correlation should increase a
lot after the progress in description of the kaon generation
process. Now we have two additional degrees of freedom ($R$ and
$\lambda$), and Fits 1--4 show that even if we only fix $\lambda$,
the data become much more "strict".

Note that Eq. (\ref{corrForm}) is not a precise formula. Here
$\lambda$ is an effective parameter that takes into account the
non-Gaussian distribution of the kaon source, etc. If the
distribution is severely non-Gaussian, Eq. (\ref{corrForm}) should
be completely modified: it is not enough to just introduce
$\lambda$. In Fits 1 and 3 and Ref. \cite{alice-2017}, the
obtained values of $\lambda$ were not close to 1 ($\approx 0.6$ in
Ref. \cite{alice-2017}). This rises the question of the
self-consistency of the results (however, it could be explained by
other effects; see Ref. \cite{lambdaClarification}). As far as we
understand, it is not easy to achieve progress in this field.

\section{Conclusion}

It was shown that the ALICE data on $K^0_SK^+$ correlation could
be described simultaneously with the Belle data on
$\gamma\gamma\to \eta\pi^0$ and the KLOE data on
$\phi\to\eta\pi^0\gamma$ in a scenario based on the four-quark
model.

Fit 2 shows that the data could be well described with the
correlation strength $\lambda$ equal to unity, as it should be for
an ideal chaotic Gaussian source. However, we emphasize that the
current experimental data does not allow us to make strict
conclusions on $\lambda$.

\section{Acknowledgements}

The present work is partially supported by the Russian Foundation
for Basic Research Grant No. 16-02-00065 and the Presidium of the
Russian Academy of Sciences Project Grant No. 0314-2015-0011.

\section{Appendix I: Polarization operators}
\label{polarizationOp}

For pseudoscalar mesons $a,b$ and $m_a\geq m_b,\ m\geq m_+$, one
has

\begin{eqnarray}
\label{polarisator}
&&\Pi^{ab}_S(m^2)=\frac{g^2_{Sab}}{16\pi}\left[\frac{m_+m_-}{\pi
m^2}\ln \frac{m_b}{m_a}+\right.\nonumber\\
&&\left.+\rho_{ab}\left(i+\frac{1}{\pi}\ln\frac{\sqrt{m^2-m_-^2}-
\sqrt{m^2-m_+^2}}{\sqrt{m^2-m_-^2}+\sqrt{m^2-m_+^2}}\right)\right]
,\end{eqnarray}

\noindent where
$\rho_{ab}(s)=2p_{ab}(s)/\sqrt{s}=\sqrt{(1-m_+^2/s)(1-m_-^2/s)}$,
and $m_\pm=m_a\pm m_b$.

\noindent For $m_-\leq m<m_+$,
\begin{eqnarray}
&&\Pi^{ab}_{S}(m^2)=\frac{g^2_{Sab}}{16\pi}\left[\frac{m_+m_-}{\pi
m^2}\ln \frac{m_b}{m_a}-|\rho_{ab}(m)|+\right.\nonumber\\
&&\left.+\frac{2}{\pi}|\rho_{ab}(m)
|\arctan\frac{\sqrt{m_+^2-m^2}}{\sqrt{m^2-m_-^2}}\right],
\end{eqnarray}
\noindent and for $m<m_-$,
\begin{eqnarray}
&&\Pi^{ab}_{S}(m^2)=\frac{g^2_{Sab}}{16\pi}\left[\frac{m_+m_-}{\pi
m^2}\ln \frac{m_b}{m_a}-\right.\nonumber\\
&&\left.-\frac{1}{\pi}\rho_{ab}(m)\ln\frac{\sqrt{m_+^2-m^2}-
\sqrt{m_-^2-m^2}}{\sqrt{m_+^2-m^2}+\sqrt{m_-^2-m^2}}\right].
\end{eqnarray}

The constants $g_{Sab}$ are related to the width as
\begin{equation}
\Gamma_S(m)=\sum_{ab} \Gamma(S\to
ab,m)=\sum_{ab}\frac{g_{Sab}^2}{16\pi m}\rho_{ab}(m). \label{GRab}
\end{equation}

\section{Appendix II: Other parameters}

For completeness, we show parameters that are not described above
in Table II. One can find all of the details in Ref.
\cite{aks-2015}.

\begin{center}
Table II. Parameters not mentioned in Table I.

\begin{tabular}{|c|c|c|c|c|c|}\hline

Fit & 1 & 2 & 3 & 4 \\ \hline

$g^{(0)}_{a_0 \gamma\gamma}\,$, $10^{-3}\,$GeV$^{-1}$ & $1.8$ &
$1.8$ & $1.8$ & $1.8$
\\ \hline

$g_{a'_0 \gamma\gamma}$, $10^{-3}\,$GeV$^{-1}$ & $8.53$ & $7.73$ &
$5.5$ & $5.5$
\\ \hline

$c_0$ & $8.8$ & $8.1$ & $10.3$ & $10.3$ \\ \hline

$c_1$, GeV$^{-2}$ & $-20.1$ & $-18.4$ & $-24.2$ & $-24.2$
\\ \hline

$c_2$, GeV$^{-4}$ & $-0.001$ & $-0.002$ & $-0.0009$ & $-0.0009$
\\ \hline

$f_{K\bar K}$, GeV$^{-1}$ & $-0.305$ & $-0.34$ & $-0.51$ & $-0.51$
\\ \hline

$f_{\pi\eta'}$, GeV$^{-1}$ & $1.0$ & $1.0$ & $27.0$ & $27.0$
\\ \hline

$\delta,^{\circ}$ & $-77.3$ & $-67.8$ & $-94.5$ & $-94.5$
\\ \hline

\end{tabular}
\end{center}

\end{document}